# Routing Approach for P2P Systems Over MANET Network

**Sofian Hamad and Taoufik Yeferny**

College of Science, Northern Border University, Arar, Saudi Arabia

**Summary**
Thanks to the great progress in mobile and wireless technologies, Internet-distributed applications like P2P file sharing are nowadays deployed over MANET (i.e., P2P mobile systems). These applications allow users to search and share diverse multimedia resources over MANET. Due the nature of MANET, P2P mobile systems brought up many new thriving challenges regarding the query routing issue. To tackle this problem, we introduce a novel context-aware query routing protocol for unstructured P2P mobile file sharing systems. Our protocol (i) locates relevant peers sharing pertinent resources for user's query; and (ii) ensures that those peers would be reached by considering different MANET constraints (e.g., query content, peer mobility, battery energy, peer load). In order to consider all these constraints for choosing the relevant peers, we are based on the technique for order preferences by similarity to ideal solution (TOPSIS). We implemented the proposed protocol and compared its routing efficiency and retrieval effectiveness with another protocol taken from the literature. Experimental results show that our scheme carries out better than the baseline protocol with respect to accuracy.

## 1. Introduction

In the last few years, mobile and wireless technologies have been achieved great progress. Mobile devices have richer functionalities and equipped with low radio range technology (e.g., Bluetooth, Wi-Fi, etc.), which allows user to store and share more resources. With the great progress of wireless networks, new types of network architectures like mobile ad hoc networks (MANETs) appear. In MANET, mobile peers can communicate without a fixed infrastructure. Each peer can communicate directly with its neighborhood. To communicate with peers outside the transmission range, messages are propagated across multiple hops in the network. Because the topology of the network may change frequently, no centralize entity can be used for message routing, thus all peers are responsible for this task. Furthermore, devices in MANETs usually have limited resources such as battery power, CPU capacity, memory and bandwidth. So, protocols and applications have to be optimized for such resource's limitations.

Internet-distributed applications such as P2P file sharing are also deployed over MANET (i.e., P2P mobile systems). These applications allow users to search and share diverse multimedia resources over MANET. Due the nature of MANET, P2P mobile systems brought up many new thriving challenges regarding the query routing issue. Hence, efficient query routing protocol for P2P mobile systems has (*i*) to locate the best peers that share pertinent resources for user's query; and (*ii*) to guarantee that those peers could be reached in such dynamic and energy-limited environment.

Query routing protocols of P2P systems on wired scenario are unsuitable for P2P mobile systems, since they do not consider several contextual information, such as peer mobility, battery energy, peer load, CPU capacity, etc. Consequently, these protocols cannot guarantee that the pertinent peers would be reached in such a dynamic and energy-limited environment. In the literature, several research works have attempted to develop context-aware query routing protocols more suitable for P2P mobile systems. The common idea of these protocols consists to send recursively the query (i.e., search message) to a selected subset of pertinent neighbors. These pertinent neighbors are chosen based on user context (e.g., query content, battery energy, peer load, peer mobility, etc.). Hence, the effectiveness and the efficiency of these query routing protocols depends on the goodness of the scoring function and the used contextual information. As a result, each of the existing protocols has its own advantages and limits. To avoid the drawbacks of existing protocols, in this paper, we introduce a novel context-aware integrated query routing protocol for P2P mobile file sharing systems. The key contributions of our proposal are: (i) selecting relevant peers (i.e., peers that share pertinent resources) according to the query content and the user profile. (ii) Ensuring that the pertinent peers would be reached by considering MANET constraints. Indeed, we considered the mobility of the relay peers to ensure the link stability between the query initiator and the peers that have the relevant documents. Moreover, we tackled the congestion and energy consumption problems by avoiding routing the query respectively to the loaded and energy-limited peers. In order to consider all these constraints for choosing the most relevant peers to route the query to, we are based on the technique for order preferences by similarity to ideal solution (TOPSIS) [1].

The rest of the paper is organized as follows. In Section 2, we present a critical overview of query routing protocols for P2P mobile systems. Section 3 thoroughly discusses our solution. In Section 4, we report the results of our experimental evaluation. Section 5 concludes and sketched issues for future work.

## 2. RELATED WORK

In the literature, there are several approaches of query routing in unstructured P2P mobile systems. Bin Tang et al [2] split the existing approaches into layered or integrated design approaches.

In layered design approach routing protocols at the application layer are operated on top of an existing MANET routing protocol at the network layer. Hence, the query routing protocol at the application layer selects the pertinent logical neighborhood to forward the query to, then, it uses an existing routing protocol at the network layer (e.g. *DSDV* [3], *DSR* [4], *AODV* [5], to name but a few.) to reach them. However, due to peer mobility, these overlay neighbors may not reflect the current underlay network, and thus may need a multi-hop route to be reached. As a result, such overlay hop required by the protocol at the application layer could result in a costly flooding-based route discovery by the multi-hop routing protocol.

Within an integrated design approach, routing protocol at the application layer is integrated to a MANET routing protocol at the network layer. In this respect, several integrated query routing protocols are proposed. They are classified as overlay-based and non-overlay-based protocols [6]. Overlay-based protocols aim to build an efficient unstructured P2P overlay that closely matches the underlying Mobile Adhoc Network. To localise pertinent resources for a given query, pure flooding or controlled flooding techniques (e.g., random walk) are used. In [7], a decentralized and dynamic topology control protocol called *TCP2P* is proposed. This protocol builds a P2P overlay closely matches the underlay network, in order to increase the fairness and provides incentive in wireless P2P file sharing applications and is energy-conserving. *E-UnP2P* protocol [8] builds an efficient overlay avoiding redundant links and transmissions while ensuring connectivity among peers.

Non-overlay-based protocols define a progressive search mechanism based on the user's context that enables routing queries to the 'best' neighbors. In order to find relevant documents, the initiator peer sends the query to its most relevant neighbors, which, in turn, forward the query to their most relevant neighbors and so on, until the query time-to-live (*TTL*) expires. To select the best neighbors, a peer is based on contextual information (i.e., remaining battery energy, signal power, neighbor velocity, neighbor's content, query content, etc.). In [9], the authors propose a P2P file sharing system over MANET based on Swarm Intelligence, referred to as P2PSI. They aim to cope with mobility problem without flooding by applying the cognitive behaviour of the real ant colonies. P2PSI adopts a hybrid push-and-pull approach. It comprises advertisement (push) and discovery (pull) processes. In the advertisement process, each file holder regularly broadcasts an advertisement message to inform its neighborhood about what files are to be shared. The discovery process locates the desired file then leaves the pheromone to help subsequent search requests. *P2PSI* avoids broadcasting the query to all neighboring peers and routes it to peers that could provide an answer. However, several important factors (i.e., battery energy, load charge, mobility, etc.) are not considered in choosing the best neighbors. Furthermore, the broadcasting technique of use to inform the neighborhood about the shared file, is costly in terms of network traffic. In [10], the authors propose a query routing protocol *Gossiping − LB* based on Gossiping [11] approach. In the proposed protocol *Gossiping-LB*, when a peer looks for forwarding a given query, it computes a forwarding probability for each of its neighbors based on it computational load, then forwards the query to neighbors having a lower load. This protocol ensures a load balancing between peers by avoiding routing more queries to saturated peers. However, it floods the query regardless its content and does not consider the mobility nor the battery energy factors. In [12], the authors had optimized the random walk unstructured content discovery protocol, which is a controlled flooding technique, to be suitable for P2P mobile systems. They designed this protocol using the G-network which is a queuing system with two types of customer, negative and positive. Then, it optimizes this protocol by the gradient descend technique based on a cost function. This function is based on the average queries response time, the average queries hit rate, and the amount of energy consumption for query distribution. Although, this protocol is efficient in terms of energy consumption and response time, it is not effective since the cost function does not consider the query string but only it is based on statistics about past queries hit.

Unlike the aforementioned protocols, we introduce a context-aware routing protocol which learns the user's behavior, in order to route the forthcoming queries to neighbors keen to provide adequate answers with regard to the query content. Furthermore, our protocol considers mobility and remaining battery energy power of relay peers, respectively, to ensure that the pertinent peers would be reached and to increase the network lifetime. It also considers the peer load factor to avoid query latency.

## 3. Proposed Protocol

Our query routing protocol is designed for unstructured P2P mobile systems. The idea underlying our proposal is to replace the query flooding method *random breadth first search (RBFS)* used to search resources in unstructured network by a context-aware query routing protocol. Indeed, in *RBFS*, the initiator peer sends the search query to *K* random neighbors, which in turn, route it with the same manner to their neighbors until a time-to-live (*TTL*) counter is decremented to 0. If a contacted peer has relevant resources, then it sends a reply message, through the reverse path of the query, to the initiator peer. Although this solution is robust, it generates an overwhelming number of messages, a lot number of irrelevant peers are contacted, and it cannot quickly locate the pertinent peers. Furthermore, query hits message may not be received by the query initiator peer due to the limitation of *MANET* environment (i.e., mobility, battery power, peer load, etc). Indeed, relay peers (i.e., peers in the reverse path of the query message) may, in the meanwhile, turn off or move out of the network at any time. To avoid these drawbacks, our context-aware routing protocol routes the query to the best *K* neighbors instead of randomly chosen ones. Indeed, our proposal aims to route the query to the most relevant neighbors, which are *(1)* more suitable to answer the query with regard to its content, *(2)* more stable to relay query hits, *(3)* less loaded to avoid congestion and (4) having more remaining battery lifetime to decrease the probability of disconnecting or partitioning the network. To rank neighboring peers according to all these context features or criteria, we are based on the technique for order preferences by similarity to ideal solution (TOPSIS). Furthermore, to compute the weights for different criteria, we used the Analytic Hierarchy Process (AHP) method.

In the following, we present the different context features considered to select the best *k* neighbors for a given query *q*. Thereafter, we present our neighbors selection algorithm.

3.1. Context features

In our protocol, we consider *(1)* the user's profile and the query content; *(2)* the peer mobility; *(3)* the battery energy of mobile peer; and *(4)* the peer load context features.

**1) User's profile and query content:** In our proposal, we consider the user's profile of neighboring peers and the query content in order to help the forwarding peer to find the most pertinent peers able to answer the query. To achieve this goal, each peer maintains a profile for each of its neighbors. The profile contains a list of the most recent past queries that the specific neighbor provided answers for. For each query it receives, the forwarding peer $p_i$ estimate the similarity *Psim* between the query to forward *q* and each neighbor $n_j$ based on its profile.

To compute *Psim*, we compare *q* to previously seen queries using the cosine similarity [13]. Thereafter, we compute an aggregate similarity of $n_j$ to *q*, as follows:

$$Psim(n_j, q) = \sum_{q_k \in Q_j} Cosine(q_k, q) \quad (1)$$

Where $Q_j$ is the set of past queries answered by $n_j$.

**2) Link stability:** In MANET environment, peers are free to move from their location at any time. Consequently, links between connected peers have a limited lifetime. Hence, link stability $l_{ij}$ between the forwarding peer $p_i$ and its neighbor $n_j$ is an important factor in the query routing process. Indeed, it is not attractive for $p_i$ to route the query *q* to the neighbor $n_j$ if the link $l_{ij}$ between them is unstable (i.e., have a short lifetime). In our proposal, we consider this factor by avoiding routing the query to neighbor peers having unstable links with the forwarding peer. To predict the lifetime of a link $l_{ij}$ between a peer $p_i$ and its neighbor $n_j$, we use the RABR network protocol [14] affinity function $a_{ij}(t)$. This function estimates at instant *t*, the time taken by $n_j$ to move out of the range of peer $p_i$. In the remainder, $Stability_{ij}(t)$ denotes the lifetime of a link $l_{ij}$ between the forwarding peer $p_i$ and its neighbor $n_j$ at time *t*.

**3) Battery energy:** The battery level of a mobile peer decreases when the peer initiates data transmission or when it forwards packets. A peer gets killed (disconnected) if the battery power finishes. Furthermore, in MANET peers need to relay their messages through other peers toward their intended destinations. Consequently, a decrease in the number of connected mobile peers could increase the probability of disconnecting or partitioning the network. To tackle this issue, our protocol avoids routing the query to neighbor peers having the least battery capacity. Indeed, the less capacity a peer has, the less lifetime it has. In the remainder, $Rengy(n_j,t)$ denotes the remaining battery energy of a given neighbor peer $n_j$ at time *t*.

**4) Peer load:** The queue utilization of neighbor peer indicates whether the line is congested or not. Indeed, if a peer is not able to deliver all the packets with the same arrival rate, then it makes a queue for storing some packets for short period of time. Thus, we define a *Load* function that asses the workload of a neighbor $n_j$ at time *t* as follows:

$$Load(n_j, t) = N/S \quad (2)$$

where *N* denotes the number of packets waiting in the queue and *S* is the maximum size of the queue. Hence, a high value of *Load* indicates that the peer $n_j$ is more loaded and the we would not keen to forward the query to.

3.2 Peers Selection Algorithm

In order to select the best *k* neighbors to forward a given query *q* to, the forwarding peer $p_i$ ranks its neighborhood according the *Psim*, *Stability*, *Rengy* and *Load* criteria. Indeed, the most relevant neighbors are the ones that maximize the *Psim*, *Stability* and *Rengy* criteria (i.e., benefit criteria) and minimize the *Load* criterion (i.e., cost criterion). To achieve this goal, we used the technique for order preferences by similarity to ideal solution (TOPSIS) combined with the Analytic Hierarchy Process (AHP) method [15] for weighting the different criteria. In the following, we present a brief description of these analytic methods and we give an illustrative example of our query routing algorithm.

**1) TOPSIS:** Technique for Order Preference by Similarity to Ideal Solution (TOPSIS) method was developed by Hwang and Yoon [1] as a method for supporting multiple-criteria decision making. Indeed, this method allows ordering all alternatives under consideration according to a set of criteria. TOPSIS considers that the best alternative would be the one that is nearest to the ideal positive solution and farthest from the ideal negative solution. A positive ideal solution maximizes the benefit criteria and minimizes the cost criteria, whereas a negative ideal solution maximizes the cost criteria and minimizes the benefit criteria. In order to rank the alternatives, TOPSIS requires as input a decision matrix $X$ with $m$ rows, each representing the alternative under consideration and $n$ columns, each representing the evaluation criterion. The structure of the matrix $X$ can be expressed as follows:

$$X = \begin{array}{c} \\ A_1 \\ A_2 \\ A_3 \\ A_m \end{array} \begin{pmatrix} C_1 & C_2 & \ldots & C_n \\ x_{11} & x_{12} & \ldots & x_{1n} \\ x_{21} & x_{22} & \ldots & x_{2n} \\ \ldots & \ldots & \ldots & \ldots \\ x_{m1} & x_{m2} & \ldots & x_{mn} \end{pmatrix}$$

Where each $x_{ij}$, $i = 1,...,m$ and $j = 1,...,n$, denotes the performance of the $i^{th}$ alternative $A_i$ with respect to the $j^{th}$ criterion $C_j$. The TOPSIS algorithm consists of the six following steps:

**Step 1**: Calculate the normalized decision matrix R:

$$R = \begin{pmatrix} r_{11} & r_{12} & \ldots & r_{1n} \\ r_{21} & r_{22} & \ldots & r_{2n} \\ \ldots & \ldots & \ldots & \ldots \\ r_{m1} & r_{m2} & \ldots & r_{mn} \end{pmatrix}$$

where $r_{ij}$ is computed as follows:

$$r_{ij} = \frac{x_{ij}}{\sqrt{\sum_{i=1}^{m} x_{ij}^2}} \quad i = 1, 2, ..., m \text{ and } j = 1, 2, ..., n \quad (3)$$

**Step 2**: Calculate the weighted normalized decision matrix V:

$$V = \begin{pmatrix} v_{11} & v_{12} & \ldots & v_{1n} \\ v_{21} & v_{22} & \ldots & v_{2n} \\ \ldots & \ldots & \ldots & \ldots \\ v_{m1} & v_{m2} & \ldots & v_{mn} \end{pmatrix}$$

where the weighted normalized value $v_{ij}$ is calculated as follows:

$$v_{ij} = r_{ij} \times w_j \quad i = 1, 2, ..., m \text{ and } j = 1, 2, ..., n \quad (4)$$

where $w_j$ is the weight of the $j^{th}$ criterion.

**Step 3**: Determine the ideal ($A^+$) and negative ideal ($A^-$) solutions.

$$A^+ = \{(\max_i v_{ij} \mid j \in C_b), (\min_i v_{ij} \mid j \in C_c)\} = \{v_j^+ \mid j = 1, 2, ..., n\} \quad (5)$$

$$A^- = \{(\min_i v_{ij} \mid j \in C_b), (\max_i v_{ij} \mid j \in C_c)\} = \{v_j^- \mid j = 1, 2, ..., n\} \quad (6)$$

where $C_b$ and $C_c$ denote the respectively set of benefit and cost criteria.

**Step 4**: Calculate the separation measures using the m-dimensional Euclidean distance. The separation measures of each alternative from the positive ideal solution and the negative ideal solution, respectively, are as follows:

$$S_i^+ = \sqrt{\sum_{j=1}^{m} (v_{ij} - v_j^+)^2}, j = 1, 2, ..., n \quad (7)$$

$$S_i^- = \sqrt{\sum_{j=1}^{m} (v_{ij} - v_j^-)^2}, j = 1, 2, ..., n \quad (8)$$

**Step 5**: Calculate the relative closeness to the ideal solution. The relative closeness of the alternative $A_i$ with respect to $A^+$ is defined as follows:

$$RC_i^+ = \frac{S_i^-}{S_i^+ + S_i^-}, i = 1, 2, ..., m \quad (9)$$

**Step 6**: Rank the alternatives in descending order using *RC*.

**2) AHP:** The Analytic Hierarchy Process (AHP) [15], introduced by Thomas Saaty is an effective method that helps the decision makers to compute the weights for the different criteria. In order to achieve this, the AHP algorithm consists on the following steps:

**Step 1**: Create a pairwise comparison matrix A. The matrix A is a $n \times n$ real matrix, where $n$ is the number of evaluation criteria considered. Each entry $a_{jk}$ of the matrix A represents the importance of the $j^{th}$ criterion relative to the $k^{th}$ criterion. The entries $a_{jk}$ and $a_{kj}$ must satisfy the following constraint:

$$a_{jk} \times a_{kj} = 1 \quad (10)$$

Obviously, $a_{jj} = 1$ for all $j$. The relative importance between two criteria is measured according to a numerical scale from 1 to 9, as shown in Table 1.

Table I: Table of relative scores.

| Value of $a_{jk}$ | Interpretation |
|---|---|
| 1 | $j$ and $k$ are equally important |
| 3 | $j$ is slightly more important than $k$ |
| 5 | $j$ is more important than $k$ |
| 7 | $j$ is strongly more important than $k$ |
| 9 | $j$ is absolutely more important than $k$ |

|  | $S_i^+$ | $S_i^-$ | $RC_i$ | Rank |
|---|---|---|---|---|
| $n_1$ | 0.055 | 0.170 | 0.756 | 1 |
| $n_2$ | 0.103 | 0.168 | 0.619 | 2 |
| $n_3$ | 0.167 | 0.072 | 0.300 | 5 |
| $n_4$ | 0.152 | 0.141 | 0.480 | 3 |
| $n_5$ | 0.149 | 0.089 | 0.373 | 4 |

**Step 2**: Create a normalized pairwise matrix $A_{norm}$ of $A$, where each entry $a_{ij}$ of the matrix $A_{norm}$ is computed as follows:

$$\bar{a}_{ij} = \frac{a_{ij}}{\sqrt{\sum_{i=1}^{n} a_{ij}}} \quad (11)$$

**Step 3**: Create the criteria weight vector $w$, which is an n-dimensional column vector. The weight vector $w$ is built by averaging the entries on each row of $A_{norm}$, as follows:

$$w_j = \frac{\sqrt{\sum_{j=1}^{n} \bar{a}_{ij}}}{n} \quad (12)$$

**3) Illustrative example:** Assume that a given peer $P$ has to select the 3 most relevant peers from a set of neighbors $N = \{n_1, n_2, n_3, n_4, n_5\}$, according to the *Psim*, *Stability*, *Rengy* and *Load* criteria. Hence, the decision matrix could be represented as follows:

$$X = \begin{array}{c} n_1 \\ n_2 \\ n_3 \\ n_4 \\ n_5 \end{array} \begin{pmatrix} Psim & Stability\,(Sec) & Rengy\,(J) & Load \\ 0.9 & 12 & 90 & 0.2 \\ 1 & 40 & 85 & 0.1 \\ 0.5 & 34 & 60 & 0.7 \\ 0.1 & 60 & 70 & 0.6 \\ 0.3 & 50 & 10 & 0.8 \end{pmatrix}$$

Before starting the execution of the TOPSIS algorithm, we will compute the weight of each criterion based the AHP method. In this respect we defined the following pairwise matrix:

$$A = \begin{array}{c} Psim \\ Stability \\ Rengy \\ Load \end{array} \begin{pmatrix} Psim & Stability & Rengy & Load \\ 1 & 1/5 & 3 & 3 \\ 5 & 1 & 5 & 3 \\ 1/3 & 1/5 & 1 & 1 \\ 1/3 & 1/3 & 1 & 1 \end{pmatrix}$$

The next step is to normalize the matrix A:

$$A_{norm} = \begin{pmatrix} 0.150 & 0.115 & 0.300 & 0.375 \\ 0.750 & 0.577 & 0.500 & 0.375 \\ 0.050 & 0.115 & 0.100 & 0.125 \\ 0.050 & 0.192 & 0.100 & 0.125 \end{pmatrix}$$

Hence, the weighing vector is:
$W = (w_{psim}, w_{stability}, w_{rengy}, w_{load}) = (0.235, 0.55, 0.098, 0.117)$

As we aforementioned, the TOPSIS algorithm consists of six steps:

**Step 1**: Calculate the normalized decision matrix R:

$$R = \begin{pmatrix} 0.612 & 0.469 & 0.582 & 0.161 \\ 0.680 & 0.375 & 0.550 & 0.081 \\ 0.340 & 0.319 & 0.388 & 0.564 \\ 0.068 & 0.563 & 0.453 & 0.483 \\ 0.204 & 0.469 & 0.065 & 0.645 \end{pmatrix}$$

**Step 2**: Calculate the weighted normalized decision matrix V:

$$V = \begin{pmatrix} 0.144 & 0.258 & 0.057 & 0.019 \\ 0.160 & 0.206 & 0.054 & 0.009 \\ 0.080 & 0.175 & 0.038 & 0.066 \\ 0.016 & 0.310 & 0.044 & 0.057 \\ 0.048 & 0.258 & 0.006 & 0.075 \end{pmatrix}$$

**Step 3**: Determine the ideal ($A^+$) and negative ideal ($A^-$) solutions. In our case, the benefit criteria are Psim, Stability, Rengy and the cost criterion is Load.
$A^+ = (0.160, 0.310, 0.057, 0.009)$
$A^- = (0.016, 0.175, 0.006, 0.075)$

**Step 4**: Table II shows the separation measures $S_i^+$ and $S_i^-$ of each neighbor, respectively from the positive ideal solution and the negative ideal solution.

Table II: Computing the Separation Measures and the Relative Closeness
**Step 5**: Table II shows the relative closeness $RC_i$ of each neighbor to the ideal solution.

**Step 6**: Table II shows the rank of each neighbor. Indeed, in this example the query will be routed to the neighboring peers $n_1$, $n_2$ and $n_4$.

## 4. Performance Evaluation

Our experimental study aims to validate the following issues, discussed in this paper:
- The selection of pertinent peers for a given query.
- Ensuring that the pertinent peers would be reached by considering MANET constraints.

To validate our protocol, we compare its routing effectiveness as well as its efficiency versus the *Gossiping-LB* protocol [10] by using the NS2 simulator. Indeed, we have chosen the *Gossiping-LB* protocol in order to compare our protocol versus another one which routes the query regardless its content. Furthermore, our vision of link stability is quite different from that of Gossiping-LB.

To evaluate the routing *effectiveness* of our protocol, we used the recall and the average hit rate metrics [15,16]. Besides, we are interested in assessing the routing *efficiency* of our protocol by computing the average file-discovery delay per query [17,18].

4.1 Results

**Routing effectiveness of *CAQRP* and *Gossiping-LB***: Figure 1 shows that our protocol, denoted in the sequel *CAQRP* (Context-Aware Query Routing Protocol), outperforms *Gossiping-LB* in terms of recall and hit rate under different network sizes. Figure 1 (b) shows that our protocol increases the hit rate of *Gossiping-LB* by 70% and 183%, respectively with 25 and 100 peers. Indeed, our protocol considers the query content which allows to better target the relevant neighbors. However, *Gossiping-LB* routes the query to a less loaded neighborhood regardless the query content. Figure 1 also shows that the hit rate and the recall decrease for both protocols whenever the the network size goes up.

**Routing efficiency of *CAQRP* and *Gossiping-LB***: Figure 2 shows that by increasing the network size, the average file discovery delay of both protocols also increases. In addition, it depicts that our protocol has shorter average-file discovery delay than *Gossiping-LB* at all variations in the overlay size. In fact, our protocol exploits the similarity between the query to route and the past ones, which help to quickly find pertinent peers reducing by the way the number of hops. Hence, there is few relays between the query initiating peer and the target peers.

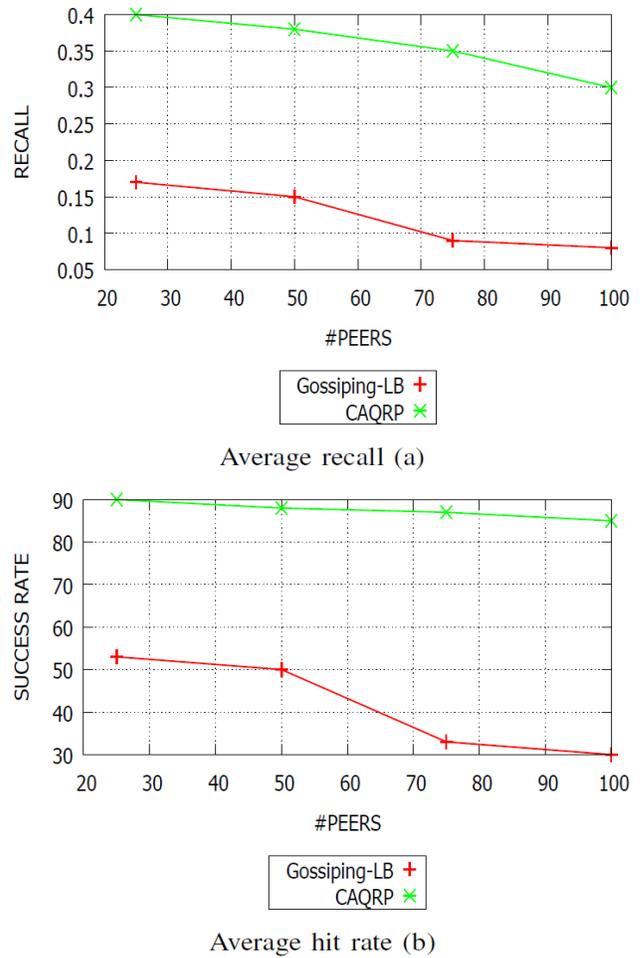

Fig. 1  Average recall and hit rate

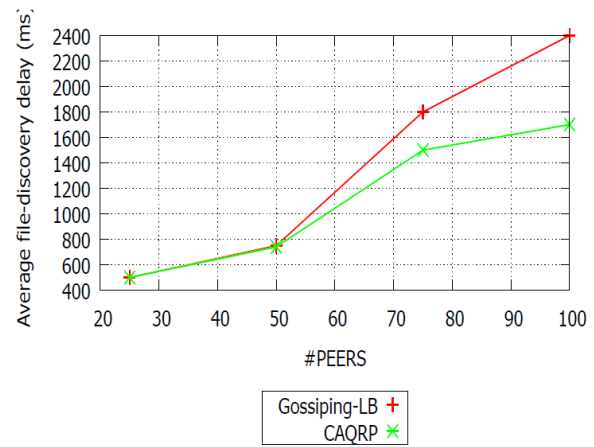

Fig. 2  Average file-discovery delay

## 5. Conclusion

In this paper, we have introduced a query routing protocol for unstructured P2P systems. Indeed, in our protocol, we have considered the mobility of the relay peers to ensure links stability between the query initiator and the peers holding relevant documents to the query. Moreover, we tackled the congestion and energy consumption problems by avoiding routing the query respectively to the loaded and energy-limited peers. In order to consider all these constraints for choosing the most relevant peers to route the query to, we have based on the technique for order preferences by similarity to ideal solution. Performed simulations highlight the effectiveness and the efficiency of our protocol. Obvious pointers for future work include an in-depth study on the impact of the weighting parameters on the efficiency of our protocol.

## Acknowledgments

This project was funded by Deanship of Scientific Research, Northern Border University for their financial support under grant no. SCI-2018-3-9-F-8026. The authors, therefore, acknowledge with thanks DSR technical and financial support.

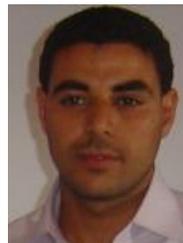

**Taoufik Yeferny** He received the M.C.S. and Ph.D. degrees in computer sciences from the University of Tunis El Manar, Tunisia, in 2009 and 2014, respectively. From 2013 to 2016, he was an Assistant Professor with the High Institute of Applied Languages and Computer Science of Beja, Tunisia. Since 2016, he has been an Assistant Professor with Northern Border University, Saudi Arabia. His current research interests include mobile P2P systems, vehicle ad hoc networks, and intelligent transportation systems.

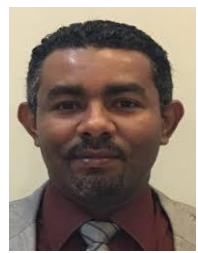

**Sofian Hamad** He received the B.Sc. degree (Hons.) in computer science from Future University, Khartoum, Sudan, in 2003, the M.Sc. degree in management business administration (MBA) from the Sudan Academy of Science, in 2007and the Ph.D. degree in electrical engineering from Brunel University, London, U.K., in 2013. He is currently an Assistant Professor with the Department of Computer Science, Northern Border University. His current research interests include ad-hoc and mesh wireless networks, vehicular technology, and the Internet of Things.